\documentclass[12pt]{article}

\normalbaselines

\def\smallromani{\renewcommand{\theenumi}{\roman{enumi}}
\renewcommand{\labelenumi}{(\theenumi)}}

\newcommand{\Proof}{\NI
                    {\bf Proof.}\ }

\usepackage{handbookbib}
\usepackage{url}

\usepackage{amsmath}
\usepackage{amsfonts}
\usepackage{amssymb}
\usepackage{latexsym}

\usepackage{enumerate}

\newcommand{\bfe}[1]{\begin{bfseries}\emph{#1}\end{bfseries}\index{#1}}
\newcommand{\oldbfe}[1]{\begin{bfseries}\emph{#1}\end{bfseries}}
\newcommand{\bfei}[1]{\begin{bfseries}\emph{#1}\end{bfseries}\index{#1}}

\newcommand{\ES}{\mbox{$\emptyset$}}

\newcommand{\myra}{\mbox{$\:\rightarrow\:$}}

\newcommand{\tra}{\mbox{$\:\rightarrow^*\:$}}

\newcommand{\sse}{\mbox{$\:\subseteq\:$}}

\newcommand{\fa}{\mbox{$\forall$}}
\newcommand{\te}{\mbox{$\exists$}}

\newcommand{\LL}{\mbox{$\ldots$}}

\newcommand{\C}[1]{\mbox{$\{{#1}\}$}}           % curly braces

\newcommand{\NI}{\noindent}
\newcommand{\HB}{\hfill{$\Box$}}
\newcommand{\VV}{\vspace{5 mm}}

\usepackage{amsfonts}
\usepackage{latexsym}
\usepackage{alltt}

\newtheorem{theorem}{Theorem}%[section]
\newtheorem{defined}{Definition}

\newtheorem{exa}{Example}

\newtheorem{lemma}{Lemma}

\usepackage{proof}

\usepackage{rotating}

 \title{Direct Proofs of Order Independence}

 \author{Krzysztof R. Apt\footnote{Centrum for
     Mathematics and Computer Science (CWI), Science Park 123, 1098 XG
     Amsterdam, the Netherlands,
     and University of Amsterdam.
 }
 }

\begin{document}

\date{}
\maketitle
%\doublespace

 \begin{abstract}
   We establish a generic result concerning order independence of a
   dominance relation on finite games. It allows us to draw conclusions
   about order independence of various dominance relations in a direct
   and simple way.
 \end{abstract}

%  \VV

%  \NI \textbf{Keywords}: order independence, hereditarity, Newman's Lemma.

%  \NI
%  \textbf{JEL Classification}: C72

%  \newpage
%  \setcounter{page}{1}

\section{Introduction}

In the literature on strategic games several dominance relations have
been considered and for various of them order independence was
established. Just to mention two well-known results.  In \cite{GKZ90}
order independence of strict dominance by a pure strategy in finite
games was established.  In turn, in \cite{OR94} order independence of
strict dominance by a mixed strategy was proved.

A number of other order independence results have been proved.  In
\cite{Apt04} we provided a uniform exposition based on so-called
abstract reduction systems, notably Newman's Lemma, and established
some new order independence results. However, for each considered
dominance relation some supplementary lemmas were needed.

The purpose of this paper is to provide a generic order independence
result for finite games that allows us to prove order independence of
each relevant dominance relation in a direct and simple way.  The
exposition still relies on Newman's Lemma, but is now directly linked
with the dominance relations through a crucial notion of hereditarity.
To check for order independence it suffices to show that the dominance
relation is hereditary, a simple condition referring to a single
reduction step. We show that in each case this is straightforward.
In the conclusions we clarify what makes this
approach simpler than the ones used in the literature.

\section{Dominance relations}  

We assume the customary notions of a \oldbfe{strategic game}, of
\oldbfe{strict dominance} and \oldbfe{weak dominance} by a pure,
respectively mixed strategy, see, e.g., \cite{OR94}.  We also use the
standard notation. In particular, $\Delta S$ is the set of
probabilities over the finite non-empty set $S$ and for a joint
strategy $s$, $s_i$ is the strategy of player $i$ and $s_{-i}$ is the
joint strategy of the opponents of player $i$.  All considered games
are assumed to be finite.

We assume an initial (finite) strategic
game
\[
G := (G_1, \LL, G_n, p_1, \LL, p_n).
\]
where $G_i$ is a non-empty set of strategies of player $i$ and
$p_i$ his payoff function. Given non-empty sets of strategies
$R_1, \ldots, R_n$ such that for all $i$, $R_i \sse G_i$ we say that $R :=
(R_1, \ldots, R_n, p_1, \ldots, p_n)$ is a \bfe{restriction} (of $G$).
Here of course we view each $p_i$ as a function on the subset $R_1
\times \ldots \times R_n$ of $G_1 \times \ldots \times G_n$.

In what follows, given a restriction $R$ we denote by $R_i$ the set of
strategies of player $i$ in $R$.  Further, given two restrictions $R$
and $R'$ we write $R' \sse R$ when for all $i$, $R'_i \sse R_i$.

When reasoning about never best responses we want to carry out the
argument for a number of alternatives in a uniform way. To this end by
a \bfe{set of beliefs} of player $i$ in a restriction $R$ we mean one
of the following sets

\begin{enumerate} \smallromani
\item ${\cal B}_{i} := R_{-i}$,

i.e., a belief is a joint pure strategy of the opponents,

\item ${\cal B}_{i} := \Pi_{j \neq i} \Delta R_{j}$,

i.e., a belief is a joint mixed strategy of the opponents,

\item 
${\cal B}_{i} := \Delta R_{-i}$,

i.e., a belief is a probability distribution over the set of joint pure
strategies of the opponents (so called \oldbfe{correlated mixed
  strategy}).
\end{enumerate}

In the second and third case the payoff function $p_i$ is extended in
a standard way to the \oldbfe{expected payoff} function $p_i : R_i
\times {\cal B}_{i} \myra \cal{R}$.

Consider now a restriction $R:= (R_1, \LL, R_n, p_1, \LL, p_n)$ with for each
player $i$ a set of beliefs ${\cal B}_{i}(R)$.  We say then that a
strategy $s_i$ in $R$ is a \oldbfe{never best response} if
\[
\fa \mu_{i} \in {\cal B}_{i}(R) \: \te s'_i \in R_i \: p_{i}(s'_i,
\mu_{i}) > p_{i}(s_i, \mu_{i}).
\]

By a \bfei{dominance relation} $D$ we mean a function that assigns to each
restriction $R$ a subset $D_R$ of $\bigcup_{i = 1}^{n} R_i$. Instead
of writing $s_i \in D_R$ we say that \bfe{$s_i$ is $D$-dominated in
  $R$}.  To avoid unnecessary complications we assume that for each
restriction $R$ and player $i$ the set of his $D$-undominated strategies in $R$ is
non-empty, i.e., that for each $i$, $R_i \setminus D_R \neq \ES$.  This
natural assumption is satisfied by all considered dominance relations.

Given two restrictions $R$ and $R'$
we write $R \myra_{\hspace{-1mm} D \: } R'$ when $R \neq R', R' \sse R$ and 
\[
\mbox{each strategy $s_i \in (\bigcup_{j = 1}^{n} R_j) \setminus
  (\bigcup_{j = 1}^{n} R'_j)$ is $D$-dominated in $R$}.
\]

An \bfe{outcome} of an iteration of $\myra_{\hspace{-1mm} D}$ starting in a game $G$
is a restriction $R$ that can be reached from $G$ using $\myra_{\hspace{-1mm} D}$ in finitely many steps
and such that for no $R'$, $R \myra_{\hspace{-1mm} D \: } R'$ holds.

We call a dominance relation $D$
\begin{itemize}

\item \bfe{order independent} 
if for all initial games $G$ all iterations of $\myra_{\hspace{-1mm} D}$ starting in $G$ yield the same
final outcome,

\item \bfe{hereditary} if for all initial games $G$, all restrictions $R$ and $R'$ 
such that $R \myra_{\hspace{-1mm} D \: } R'$ and a
strategy $s_i$ in $R'$
\[
\mbox{$s_i$ is $D$-dominated in $R$  implies that $s_i$ is $D$-dominated in $R'$,}
\]

\item \bfe{monotonic} if for all initial games $G$, all restrictions $R$ and $R'$ 
such that $R' \sse R$ and a
strategy $s_i$ in $R'$
\[
\mbox{$s_i$ is $D$-dominated in $R$  implies that $s_i$ is $D$-dominated in $R'$.}
\]

% \item \bfe{smooth} if for all initial games $G$, all restrictions $R, R'$ and $R''$ 
% such that $R \myra_{\hspace{-1mm} D \: } R' \myra^{*}_{\hspace{-1mm} D \: } R''$ and a
% strategy $s_i$ in $R''$
% \[
% \mbox{$s_i$ is $D$-dominated in $R''$ and in $R$
% implies that $s_i$ is $D$-dominated in $R'$.}
% \]
\end{itemize}

Clearly every monotonic dominance relation is hereditary, but the converse does not need to hold.

\section{Proofs of order independence}

We shall establish the following result.

\begin{theorem} \label{thm:1}
  Every hereditary dominance relation $D$ is order independent.
\end{theorem}

% The converse holds under the assumption of smoothness.

% \begin{theorem}\label{thm:2}
%   Every order independent and smooth dominance relation $D$ is hereditary.
% \end{theorem}

We relegate the proof in the appendix.  This theorem can be used to
prove in a straightforward way order independence of various
dominance relations.  We illustrate it now by means of various
examples.

\subsection{Strict dominance}

Recall that a strategy $s_i$ is \oldbfe{strictly dominated} in a restriction
$(R_1, \LL, R_n,$  $p_1, \LL, p_n)$ if for some strategy $s'_i \in R_i$
\[
\fa s_{-i} \in R_{-i} \: p_{i}(s'_i, s_{-i}) > p_{i}(s_i, s_{-i}).
\]

Denote the reduction relation corresponding to strict dominance by a
pure strategy by $\myra_{\hspace{-1mm} S}$.  To see that strict
dominance by a pure strategy is hereditary suppose that $R
\myra_{\hspace{-1mm} S \: } R'$ and that a strategy $s_i$ in $R'$ is
strictly dominated in $R$.  The initial game is finite, so there
exists in $R$ a strategy $s'_i$ that strictly dominates $s_i$ in $R$
and is not strictly dominated in $R$. Then $s'_i$ is not eliminated in
the step $R \myra_{\hspace{-1mm} S \: } R'$ and hence is a strategy in
$R'$.  But $R' \sse R$, so $s'_i$ also strictly dominates $s_i$ in
$R'$.

In contrast, strict dominance is not monotonic for the simple reason
that no strategy is strictly dominated in a game in which each player has
exactly one strategy.  Order independence of strict dominance for
finite games was originally proved in \cite{GKZ90}, by relying on the
notion of monotonicity (called there hereditarity) used for binary
dominance relations. This approach works since
a slightly different reduction relation is used there,
% $R \myra_{\hspace{-1mm} S' \: } R'$, defined by 
% \[
% \mbox{each $s_i \in (\bigcup_{j = 1}^{n} R_j) \setminus
%   (\bigcup_{j = 1}^{n} R'_j)$ is strictly dominated in $R$ by some $s'_i \in R'_i$}.
% \]
according to which every eliminated strategy is strictly dominated in
the original restriction by a strategy that is \emph{not} eliminated.
It is straightforward to check that this reduction relation coincides
with $\myra_{\hspace{-1mm} S}$.

\subsection{Global strict dominance}

We say that a strategy $s_i$ is \oldbfe{globally strictly dominated} in a restriction
$(R_1, \LL, R_n,$ $p_1, \LL, p_n)$ if for some strategy $s'_i \in G_i$ (so \emph{not} $s'_i \in R_i$)
\[
\fa s_{-i} \in R_{-i} \: p_{i}(s'_i, s_{-i}) > p_{i}(s_i, s_{-i}).
\]

This notion of dominance was originally considered in \cite[ pages
1264-1265]{MR90}. Its order independence for finite games is a
consequence of a more general result proved in \cite[ pages
200-201]{Rit02} and also \cite{CLL07}, where this dominance relation
was analyzed for arbitrary games.

Global strict dominance is clearly monotonic, so it is hereditary.

\subsection{Never best response}

Suppose that each player $i$ in the initial game $G$ has a set of
beliefs ${\cal B}_{i}$.  Then in each restriction $R$ we choose the
corresponding set of beliefs ${\cal B}_{i}(R)$ of player $i$.  For
instance, if ${\cal B}_{i} = \Pi_{j \neq i} \Delta G_{j}$ then we set
${\cal B}_{i}(R) := \Pi_{j \neq i} \Delta R_{j}$.

Note that if $R' \sse R$, then we can identify $\Pi_{j \neq i} \Delta
R'_{j}$ with a subset of $\Pi_{j \neq i} \Delta R_{j}$ and $\Delta
R'_{-i}$ with a subset of $\Delta R_{-i}$.  So we can assume that
$R' \sse R$ implies that ${\cal B}_{i}(R') \sse {\cal B}_{i}(R)$.

To prove that being a never best response is a hereditary dominance
relation consider the corresponding reduction relation between
restrictions that we denote by $\myra_{\hspace{-1mm} N}$.  Suppose
that $R \myra_{\hspace{-1mm} N \: } R'$ and that a strategy $s_i$ in
$R'$ is a never best response in $R$.  Assume by contradiction that
for some $\mu_{i} \in {\cal B}_i(R')$, $s_i$ is a best response to
$\mu_{i}$ in $R'$, i.e.,
\[
\fa s'_i \in R'_i \: p_i(s_i, \mu_{i}) \geq p_i(s'_i, \mu_{i}).
\]

We have $\mu_{i} \in {\cal B}_i(R)$
since ${\cal B}_{i}(R') \sse {\cal B}_{i}(R)$.
Take a best response $s'_i$ to $\mu_{i}$ in $R$.
Then $s'_i$ is not eliminated in
the step $R \myra_{\hspace{-1mm} N \: } R'$ and hence is a strategy in
$R'$. But by the choice of $s_i$ and $s'_i$
\[
p_i(s'_i, \mu_{i}) > p_i(s_i, \mu_{i}),
\]
so we reached a contradiction.

Order independence of iterated elimination of never best responses was
originally proved in \cite{Apt05} by comparing it with the iterated
elimination of global never best responses, a notion we discuss next.

\subsection{Global never best response}

Suppose again that each player $i$ in the initial game $G$ has a set of
beliefs ${\cal B}_{i}$. Choose in each restriction $R$ the
corresponding set of beliefs ${\cal B}_{i}(R)$ of player $i$.  

We say that a strategy $s_i$ in a restriction $R$ is a \oldbfe{global
  never best response} if

\[
\fa \mu_{i} \in {\cal B}_{i}(R) \: \te s'_i \in G_i \: p_{i}(s'_i, \mu_{i}) > p_{i}(s_i, \mu_{i}).
\]

So in defining a global never best response we compare the given
strategy with all strategies in the \emph{initial} game and not the
\emph{current} restriction.  Note that iterated elimination of global
never best responses, when performed `at full speed' yields the set
of \oldbfe{rationalizable strategies} as defined in \cite{Ber84}.

The property of being a global never best response is clearly
monotonic, so it is hereditary.  Order independence of this dominance
relation was originally proved in \cite{Apt05} for arbitrary, so
possibly infinite, games.

\subsection{Strict dominance by a mixed strategy}

% We say that a strategy $s_i$ is \oldbfe{strictly dominated by a mixed strategy} in a game 
% $(G_1, \LL, G_n, p_1, \LL, p_n)$ if for some mixed strategy $m_i$
% if
% \[
% \fa s_{-i} \in G_{-i} \: p_{i}(m_i, s_{-i}) > p_{i}(s_i, s_{-i}).
% \]
Denote the corresponding reduction relation between restrictions 
by $\myra_{\hspace{-1mm} SM}$.
Given two mixed strategies $m_{i}, m'_{i}$ and a strategy $s_i$ we
denote by $m_{i}[s_i/m'_i]$ the mixed strategy obtained from $m_{i}$ by
substituting the strategy $s_i$ by $m'_i$ and by `normalizing' the
resulting sum. 
%
% We also use the following identification of mixed strategies over two
% sets of strategies $S'_i$ and $S_i$ such that $S'_i \sse S_i$.  We
% view a mixed strategy $m_i \in \Delta S_i$ such that $support(m_i)
% \sse S'_i$ as a mixed strategy `over' the set $S'_i$, i.e., as an
% element of $\Delta S'_i$, by limiting the domain of $m_i$ to $S'_i$.
% Further, we view each mixed strategy $m_i
% \in \Delta S'_i$ as a mixed strategy `over' the set $S_i$, i.e., as an
% element of $\Delta S_i$, by assigning the probability 0 to the
% elements in $S_i \setminus S'_i$.  
First, we establish the following auxiliary lemma. 

\begin{lemma}[Persistence]
\label{lem:persist}
Assume that $R \myra_{\hspace{-1mm} SM} R'$ and that
a strategy $s_i$ in $R$ is strictly dominated in $R$ by a
mixed strategy from $R$. Then $s_i$ is strictly dominated in $R$ by a mixed
strategy from $R'$. 

\end{lemma}

\Proof
We shall use the following obvious properties of strict dominance by a mixed
strategy in a given restriction:

\begin{enumerate}[(a)]
\item for all $\alpha \in (0,1]$, 
if $s_i$ is strictly dominated by $(1 - \alpha) s_i + \alpha \: m_i$, then
$s_i$ is strictly dominated by $m_i$,

\item if $s_i$ is strictly dominated by $m_i$ and $s'_i$ is strictly dominated by $m'_i$, then
$s_i$ is strictly dominated by $m_{i}[s'_{i}/m'_i]$.

\end{enumerate}

Suppose that $R_i \setminus R'_i = \{t_{i}^1, \LL, t_{i}^k\}$.  By
definition for all $j \in \{1, \LL, k\}$ there exists in $R$ a mixed strategy $m^{j}_i$
such that $t_{i}^j$ is strictly dominated  in $R$ by $m^{j}_i$.
We first prove by complete induction that for all $j \in \{1, \LL, k\}$ there exists in $R$ 
a mixed strategy $n^{j}_i$ such that 
\begin{equation}
  \label{equ:dots}
\mbox{$t_{i}^j$ is strictly dominated in $R$ by $n^{j}_i$ and 
$support(n^{j}_i) \cap \{t_{i}^1, \LL, t_{i}^{j}\} = \ES$.}    
\end{equation}

For some $\alpha \in (0,1]$ and a mixed strategy $n_{i}^{1}$ with $t_{i}^1 \not
\in support(n_{i}^{1})$ we have
\[
m^{1}_{i} = (1 - \alpha) t_{i}^1 + \alpha \: n_{i}^{1}.
\]
By assumption $t_{i}^1$ is strictly dominated in $R$ by $m^{1}_{i}$, so by property (a) 
$t_{i}^1$ is strictly dominated in $R$ by $n_{i}^{1}$,
which proves (\ref{equ:dots}) for $j = 1$.

Assume now that $\ell < k$ and that (\ref{equ:dots}) holds for all $j \in \{1, \LL, \ell\}$.  
By assumption $t_{i}^{\ell+1}$ is strictly dominated in $R$ by $m_{i}^{\ell+1}$. 

Let
\[
m''_i := m^{\ell+1}_i [t_{i}^1/n_{i}^{1}] \LL [t_{i}^\ell/n^{\ell}_i]. 
\]
By the induction hypothesis and property (b) $t^{\ell+1}_i$ is strictly dominated in $R$ by $m''_i$ and 
$support(m''_i) \cap \C{t_{i}^1, \LL, t_{i}^{\ell}} = \ES$.

For some $\alpha \in (0,1]$ and a mixed strategy $n_{i}^{\ell+1}$ with $t_{i}^{\ell+1} \not
\in support(n_{i}^{\ell+1})$ we have
\[
m''_{i} = (1 - \alpha) t_{i}^{\ell+1} + \alpha \: n_{i}^{\ell+1}.
\]
By (a) 
$t_{i}^{\ell+1}$ is strictly dominated in $R$ by $n_{i}^{\ell+1}$. 
Also
$support(n_{i}^{\ell+1}) \cap \C{t_{i}^1, \LL, t_{i}^{\ell+1}} = \ES$,
which proves (\ref{equ:dots}) for $j = \ell+1$.

Suppose now that the strategy $s_i$ is strictly dominated in $R$ by a
mixed strategy $m_i$ from $R$. Define 
\[
m'_i := m_{i} [t_{i}^1/n_{i}^{1}] \LL [t_{i}^k/n^{k}_i].
\]
Then by property (b) and (\ref{equ:dots}) $s_i$ is strictly dominated in $R$ by
$m'_i$ and $support(m'_i) \sse R'_i$, i.e., $m'_i$ is a mixed
strategy in $R'$. 
\HB 
\VV

Hereditarity of $\myra_{\hspace{-1mm} SM}$  is now an immediate
consequence of the Persistence Lemma \ref{lem:persist}. Indeed,
suppose that $R \myra_{\hspace{-1mm} SM} R'$ and
that $s_i \in R'_i$ is strictly dominated in $R$ by
a mixed strategy in $R$. By the Persistence Lemma \ref{lem:persist}
$s_i$ is strictly dominated in $R$ by a mixed strategy in $R'$. So
$s_i$ is also strictly dominated in $R'$ by a mixed strategy in $R'$.

The proof of order independence of strict dominance by a mixed
strategy due to \cite[ pages 61-62]{OR94} relied on the existence
of Nash equilibrium in strictly competitive games.

% The Persistence Lemma also holds for the reduction relation
% $\myra_{\hspace{-1mm} WM}$ corresponding to weak dominance by a mixed
% strategy, which is not order independent.  The above argument for
% hereditarity breaks down at the last step.  Namely, if $R
% \myra_{\hspace{-1mm} WM} R'$ and a strategy $s_i$ in $R'$ is weakly
% dominated in $R$ by a mixed strategy in $R'$, then we cannot conclude
% that $s_i$ is also weakly dominated in $R'$ by a mixed strategy in
% $R'$.

\subsection{Global strict dominance by a mixed strategy}

We say that a strategy $s_i$ is \oldbfe{globally strictly dominated by
  a mixed strategy} in a restriction $(R_1, \LL, R_n, p_1, \LL, p_n)$ if for
some mixed strategy $m'_i$ in $G_i$ (so \emph{not} $m'_i$ in $R_i$)
\[
\fa s_{-i} \in R_{-i} \: p_{i}(m'_i, s_{-i}) > p_{i}(s_i, s_{-i}).
\]

This notion of dominance was studied in \cite{BFK06} (it is their
operator $\Phi$) and in \cite{Apt07}.  It is obviously monotonic
and hence hereditary.

The proof of order independence of this relation is implicit in
\cite{Apt07}.  Dominance relations are viewed there as operators on
the set of all restrictions. Its Theorem 1 states that monotonic
operators are order independent. Monotonicity property of the operator
corresponding to global strict dominance by a mixed strategy is
noted there in Section 10.

\subsection{Inherent dominance}

Consider a restriction $(R_1, \LL, R_n, p_1, \LL, p_n)$.  We say that a
strategy $s_i$ is \oldbfe{dominated given $\tilde{R}_{-i} \sse
  R_{-i}$}, where $\tilde{R}_{-i}$ is non-empty, if $s_i$ is weakly
dominated in the restriction $(R_i, \tilde{R}_{-i}, p_1, \LL, p_n)$.  Then we
say that a strategy $s_i$ is \oldbfe{inherently dominated} if for
every non-empty subset $\tilde{R}_{-i}$ of $R_{-i}$ it is weakly
dominated given $\tilde{R}_{-i}$.

This notion of dominance was introduced in \cite{Bor90}, where its
order independence was proved by establishing a connection between
inherent dominance and rationalizability. In \cite{Bor93} parts of
\cite{Bor90} were published, but not the proof of order independence.
Denote by $\myra_{\hspace{-1mm} I \: }$ the corresponding reduction
relation.

To prove that inherent dominance is hereditary suppose that $R
\myra_{\hspace{-1mm} I \: } R'$ and that a strategy $s_i$ in $R'$ is
inherently dominated in $R$.  Fix a non-empty subset $\tilde{R}_{-i}$
of $R'_{-i}$.

The initial game is finite, so there exists in $R_i$ a strategy $s'_i$
that weakly dominates $s_i$ in $(R_i, \tilde{R}_{-i}, p_1, \LL, p_n)$
and is not weakly dominated in the restriction $(R_i, \tilde{R}_{-i}, p_1,
\LL, p_n)$.  Then $s'_i$ is not eliminated in the step $R
\myra_{\hspace{-1mm} I \: } R'$ and hence is a strategy in $R'_i$.  So
$s_i$ is weakly dominated in $(R'_i, \tilde{R}_{-i}, p_1, \LL,
p_n)$ by  $s'_i$.  This proves hereditarity.

\section{Conclusions}

We established here several order independence results. They were all
proved by just checking a single property of the dominance relation,
namely hereditarity.  This approach works because of a combination of
factors.  First, as in \cite{Apt04}, we used abstract reduction
systems.  This allowed us to decouple one part
of the argument from the study of the actual dominance relations.

Second, we viewed the dominance relations as \emph{unary} relations,
whereas the common approach in the literature is to view them as
\emph{binary} relations.  Finally, we relied on the notion of
hereditarity that is weaker than monotonicity.  These changes allowed
us to treat various forms of strict dominance and of being a never
response, and inherent dominance in a uniform way.

We conclude by offering the following observation.  It is clear how to
define the intersection of dominance relations.  The intersection of
order independent dominance relations does not need to be order
independent.  On the other hand, the intersection of hereditary
dominance relations is clearly hereditary.  So our approach also
allows us to draw conclusions about order independence of
intersections of the discussed dominance relations.

\bibliography{/ufs/apt/bib/e,/ufs/apt/bib/apt}
\bibliographystyle{handbk}

\section*{Appendix}

We present here the proof of Theorem \ref{thm:1}.
As in \cite{Apt04} we shall use the notion of an \bfei{abstract
  reduction system}, extensively studied in \cite{Ter03}.  It is
simply a pair $(A,\myra)$ where $A$ is a set and $\myra$ is a binary
relation on $A$.  Let $\tra$ denote the transitive reflexive closure
of $\myra$.  So in particular, if $a = b$, then $a \tra b$.

We say that $b$ is a $\myra$-\bfe{normal form of} $a$
if $a \tra b$ and no $c$ exists such that $b \myra c$,
and omit the reference to $\myra$ if it is clear from the context.
If every element of $A$ has a unique normal form, we say that
$(A,\myra)$ (or just $\myra$ if $A$ is clear from the context)
satisfies the \bfei{unique normal form property}.

We say that $\myra$ is \bfe{weakly confluent}
if for all $a,b,c \in A$ 

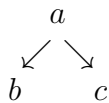
\begin{center}
$a$                                             \\
$\swarrow$ $\searrow$                           \\
$b$\ \ \ \ \ \ \ $c$                            \\
\end{center}

\NI
implies that for some $d \in A$

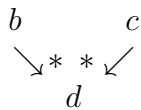
\begin{center}
$b$\ \ \ \ \ \ \ \ \ \ $c$                        \\
$\searrow \!*$ \ $\!*\swarrow$                    \\
$d$

\end{center}

The following crucial lemma is due to \cite{New42}.

\begin{lemma} [Newman] \label{lem:newman}
Consider an abstract reduction system $(A,\myra)$ such that

\begin{itemize}
\item no infinite $\myra$ sequences exist,

\item $\myra$ is weakly confluent.

\end{itemize}
Then $\myra$ satisfies the unique normal form property.
\end{lemma}

\Proof 
By the first assumption every element of $A$ has a normal
form.  To prove uniqueness call an element $a$ \emph{ambiguous} if it
has at least two different normal forms.  We show that for
every ambiguous $a$ some ambiguous $b$ exists such that $a \myra b$.
This proves absence of ambiguous elements by the first assumption.

So suppose that some element $a$ has two distinct normal forms $n_1$
and $n_2$.  Then for some $b, c$ we have $a \myra b \tra n_1$ and $a
\myra c \tra n_2$.  By weak confluence some $d$ exists such that $b
\tra d$ and $c \tra d$. Let $n_3$ be a normal form of $d$. It is also
a normal form of $b$ and of $c$. Moreover $n_3 \neq n_1$ or $n_3 \neq
n_2$. If $n_3 \neq n_1$, then $b$ is ambiguous and $a \myra b$.  
And if $n_3 \neq n_2$, then $c$ is ambiguous and $a \myra c$.  
\HB
\VV

Clearly, order independence of a dominance relation $D$ is equivalent
to the statement that for all initial games $G$ the reduction relation
$\myra_{\hspace{-1mm} D}$ satisfies the unique normal form property on
the set of all restrictions.

\VV

\NI
\textbf{Proof of Theorem \ref{thm:1}}.

Consider a restriction $R$.  Suppose that $R \myra_{\hspace{-1mm} D \:
  } R'$ for some restriction $R'$.  Let $R''$ be the restriction of
$R$ obtained by removing all strategies that are $D$-dominated in $R$.

We have $R'' \sse R'$. Assume that $R' \neq R''$. 
Choose an arbitrary strategy $s_i$ such
that $s_i \in R'_i \setminus R''_i$.  So $s_i$ is $D$-dominated in
$R$. By the hereditarity of $D$, $s_i$ is also $D$-dominated in $R'$.  This
shows that $R' \myra_{\hspace{-1mm} D \: } R''$.

So we proved that either $R' = R''$ or $R' \myra_{\hspace{-1mm} D
  \: } R''$, i.e., that $R' \myra^{*}_{\hspace{-1mm} D \: } R''$.
This implies that $\myra_{\hspace{-1mm} D}$ is weakly confluent.  
It suffices now to apply Newman's Lemma \ref{lem:newman}.
\HB
% \III

% \NI
% \textbf{Proof of Theorem \ref{thm:2}}.

% Suppose $D$ is not hereditary. Then for some restrictions $R$
% and $R^1$ and $s_i$ in $R^1$

% \begin{itemize}
% \item  $s_i$ is $D$-dominated in $R$,

% \item $R \myra_{\hspace{-1mm} D \: } R^1$,
  
% \item $s_i$ is not $D$-dominated in $R^1$.
% \end{itemize}

% Let $R^{2}$ be the restriction of $R$ obtained by removing all
% strategies that are $D$-dominated in $R$.
% We have $R \myra_{\hspace{-1mm} D \: } R^{2}$,
% so by the order independence
% for some restriction $R^{3}$ both $R^{1} \myra^{*}_{\hspace{-1mm} D \:
%   } R^{3}$ and $R^{2} \myra^{*}_{\hspace{-1mm} D \: } R^{3}$.
% Then $R^{3} \sse R^{2}$, so $s_i$ is not a strategy in $R^{3}$.  Since
% $s_i$ is a strategy in $R^{1}$, for some restrictions $R^4$ and $R^5$

% \begin{itemize}
% \item $s_i$ is a strategy in $R^4$,

% \item $s_i$ is not a strategy in $R^5$,

% \item $R^{1} \myra^{*}_{\hspace{-1mm} D \: } R^4 \myra_{\hspace{-1mm} D \: } R^{5} \myra^{*}_{\hspace{-1mm} D \: } R^3$.
% \end{itemize}

% So $s_i$ is $D$-dominated in $R^4$. 
% Since $R \myra_{\hspace{-1mm} D \: } R^1 \myra^{*}_{\hspace{-1mm} D \: } R^4$
% we get a contradiction with the assumption that $D$ is smooth.
% \HB

\end{document}